\documentclass[twocolumn]{aastex631}
\usepackage{amsmath}
\usepackage[T1]{fontenc}

\begin{document}

\title{Coagulation-Fragmentation Equilibrium for Charged Dust: \\ Abundance of Submicron Grains Increases Dramatically in Protoplanetary Disks}

\correspondingauthor{Vitaly Akimkin}
\email{akimkin@inasan.ru}

\author[0000-0002-4324-3809]{Vitaly Akimkin}
\affiliation{Institute of Astronomy, Russian Academy of Sciences, Pyatnitskaya str. 48, Moscow, 119017, Russia}

\author[0000-0002-1590-1018]{Alexei V. Ivlev}
\affiliation{Max-Planck Institute for Extraterrestrial Physics, Garching by Munich, D-85748, Germany}

\author[0000-0003-1481-7911]{Paola Caselli}
\affiliation{Max-Planck Institute for Extraterrestrial Physics, Garching by Munich, D-85748, Germany}

\author[0000-0003-1613-6263]{Munan Gong}
\affiliation{Max-Planck Institute for Extraterrestrial Physics, Garching by Munich, D-85748, Germany}

\author[0000-0003-1572-0505]{Kedron Silsbee}
\affiliation{University of Texas at El Paso, El Paso, TX, 79968, USA}

\begin{abstract}
Dust coagulation in protoplanetary disks is not straightforward and is subject to several slow-down mechanisms, such as bouncing, fragmentation and radial drift to the star. Furthermore, dust grains in UV-shielded disk regions are negatively charged due to collisions with the surrounding electrons and ions, which leads to their electrostatic repulsion. For typical disk conditions, the relative velocities between micron-size grains are small and their collisions are strongly affected by the repulsion. On the other hand, collisions between pebble-size grains can be too energetic, leading to grain fragmentation. The aim of the present paper is to study a combined effect of the electrostatic and fragmentation barriers on dust evolution. We numerically solve the Smoluchowski coagulation-fragmentation equation for grains whose charging occurs under conditions typical for the inner disk regions, where thermal ionization operates. We find that dust fragmentation efficiently resupplies the population of small grains under the electrostatic barrier. As a result, the equilibrium abundance of sub-micron grains is enhanced by several orders of magnitude compared to the case of neutral dust. For some conditions with fragmentation velocities $\sim1$\,m\,s$^{-1}$,  macroscopic grains are completely destroyed.

\end{abstract}

\section{Introduction} \label{sec:intro}
 Dust coagulation from sub-micron to centimeter-size range is the first and key step on the stairs of the bottom-up planet formation scenario \citep{2022arXiv220309759D}. The top-down planet formation via the gravitational instability \citep{1997Sci...276.1836B, 2010MNRAS.408L..36N} is affected by the grain size distribution as well, as dust opacity controls the temperature. Thus, understanding the microphysics of grain collisions provides a foundation for both branches of the planet formation theory.

   The rate and outcome of grain collisions are defined by their relative velocities. For small grains ($\lesssim 1\,\mu$m), the Brownian motion dominates. When (and if) dust grows further, the turbulence-induced velocities concur and pave the way for dust coagulation to the pebble-size range \citep{2016SSRv..205...41B}. However, the road from sub-micron grains to pebbles has at least two barriers hindering or even blocking the dust growth, the electrostatic and the fragmentation barriers. The first barrier operates in the micron-size range, where the grain collisions are not energetic enough to overcome the Coulomb repulsion \citep{2009ApJ...698.1122O, 2020ApJ...889...64A}.  The second barrier occurs due to too energetic collisions of pebble-size grains, which result in fragmentation rather than sticking \citep{2008A&A...480..859B}. While the mitigation of the electrostatic barrier requires high turbulence-induced collision velocities, mitigation the fragmentation barrier requires the opposite \citep{2011ApJ...731...96O}, and it is currently unclear how these two barriers can be simultaneously overcome. 

   Grain charge in astrophysical environments is defined by a variety of positive and negative charging currents, such as collisions with electrons and ions, photoelectric emission, and triboelectric charging (see short summary in \citet{2004ASPC..309..453W}). Photoelectric emission due to the ultraviolet radiation in the disk upper layers leads to predominantly positive grain charges there \citep{2011ApJ...740...77P,2015ARep...59..747A}. In UV-shielded regions, the plasma charging dominates and grains become negatively charged on average \citep{1987ApJ...320..803D}. The acquired charge is expected to be high enough for the total electrostatic prevention of collisions between micron-size grains \citep{2009ApJ...698.1122O}. In the absence of efficient gas ionization, the average dust charge can stay near zero and both positive and negative grains are present in equal amounts. In this case the Coulomb interaction may instead facilitate dust growth. This channel was shown to be effective for the triboelectrically charged millimeter size grains in very deep regions of protoplanetary disks shielded from the cosmic rays \citep{2020NatPh..16..225S}. If gas ionization cannot be neglected, the plasma charging dominates and the overall effect of grain charge is detrimental to the dust growth.

   Despite the apparent importance of the grain charge among the basic physical factors affecting dust evolution, this topic is currently weakly explored. A huge amount of work is done on self-consistent modelling of coagulation and fragmentation for neutral dust \citep{2008A&A...480..859B,2016ApJ...818..200E,2021MNRAS.501.4298L,2022ApJ...935...35S,2023MNRAS.518.3326L}. Concerning charged dust evolution, the focus so far has been on modeling the effect of electrostatic interactions on coagulation \citep{2011ApJ...731...95O, 2020ApJ...889...64A,2020ApJ...897..182X}, completely omitting the role of fragmentation. In this paper, we propose a self-consistent approach to model coagulation and fragmentation of charged dust, which allows us to study the interplay between the electrostatic and fragmentation barriers.

\section{Model}
To evaluate the effect of dust charge on its evolution, we solve numerically the Smoluchowski coagulation-fragmentation equation locally for a set of physical conditions. The electrostatic interaction affects at least two factors important for the outcome of grain collision: the collision cross section and impact velocity. Both factors are primarily controlled by long-range charge-charge (Coulomb) interaction, while short-range (induced dipole) interaction is only important in specific cases (see Appendices \ref{sec:inddipcros} and \ref{sec:vimp}). The collision cross section deviates from the geometrical one depending on the ratio between electrostatic and collision energies of two grains. Normally, in the UV-shielded regions of protoplanetary disks, the plasma charging dominates over other charging currents and grains acquire same-sign negative charges and thus repel each other. For typical strengths of turbulence-induced velocities in protoplanetary disks (corresponding to the Shakura--Sunyaev turbulence parameter $\alpha_{\rm t}\lesssim10^{-3}$), the Coulomb repulsion overcomes the kinetic energy for the collisions of small grains ($\lesssim1\,\mu$m). If the collision is electrostatically 'allowed', as in the case of small-grown and grown-grown grain collisions, the second factor emerges. The actual impact velocity $v_{\rm imp}$ at the grain contact changes in the comparison with the relative velocity defined at infinity $v_{\rm rel}$. While the Coulomb repulsion decelerate approaching grains, at small distances, the repulsion switches to the attraction due to induced dipole interaction. Depending on conditions, the impact velocity can be either lower or higher than the initial relative velocity $v_{\rm rel}$ at infinity.

The Smoluchowski coagulation-fragmentation equation for the mass distribution $f(m)$,\,cm$^{-3}$\,g$^{-1}$ (or the size distribution $f(a)$,\,cm$^{-4}$) can be rewritten in a discrete form. This leads to a system of $n$ nonlinear ordinary differential equations \citep{2009PhDT.......360B}:
\begin{equation}
\begin{aligned}
    \frac{{\rm d}N_k}{{\rm d} t} &=\frac{1}{2}\sum_{i=1}^n\sum_{j=1}^n N_i N_j K_{ij} \left[p_{ij}C_{ijk}+(1-p_{ij})F_{ijk} \right] \\
    &- N_k\sum_{i=1}^n N_i K_{ik},\label{eq:SCFE}
\end{aligned}
\end{equation}
where $N_k=f(m_k)\Delta m_k$ is the number density of grains within the grain mass bin $k$. Upon each collision between grains with masses $m_i$ and $m_j$, they undergo coagulation with probability $p_{ij}$ and fragmentation with probability $(1-p_{ij})$. Coagulation increases the grain number density in $k$-th bins around the mass $m_i+m_j$. Fragmentation produces a power-law spectrum of fragments and thus repopulates a range of mass bins with $k\le\max\{i,j\}$. The coefficients $C_{ijk}$ and $F_{ijk}$ are the mass weights defining how the total mass of colliding grains $m_i+m_j$ is redistributed to the $k$-th bin after their coagulation and fragmentation, respectively (see \citet{2009PhDT.......360B} and \citet{2011PhDT.......279B} for more details). The rate of collisions is determined by the coagulation kernel $K_{ij}$, cm$^{-3}$\,s$^{-1}$ defined as the product of relative velocity at infinity $v_{\rm rel}$ and the collision cross section for grains with radii $a_i$ and $a_j$:
\begin{equation}
    K_{ij} = W_{ij}\pi(a_i+a_j)^2v_{\rm rel}(a_i,a_j).
\end{equation}
Here the collision efficiency factor $W_{ij}$ accounts for the electrostatic interaction, including both the charge-charge (Coulomb) term and charge-dipole term. The Coulomb term can be accounted for in a simple analytical form:
\begin{equation}\label{eq:WC}
    W_0 = \max\left\{1 - \frac{U_{\rm C}}{E_{\rm kin}},0\right\},
\end{equation}
where $U_{\rm C}=Q_iQ_j/(a_i+a_j)$ is the Coulomb repulsion energy between grains with charges $Q_i$ and $Q_j$ at the point of their contact, and $E_{\rm kin}=\mu_{ij}v_{\rm rel}^2/2$ is their initial kinetic energy, $\mu_{ij}=m_im_j/(m_i+m_j)$ is the reduced mass of grains. The induced dipole term can be taken into account numerically (see Appendix \ref{sec:inddipcros}). Analysis shows that this term affects the collisional cross section in a narrow range of grain sizes and has limited impact on the resulting size distribution. Therefore, we include only Coulomb contribution to the collision efficiency factor, i.e. we take $W_{ij}=W_0$.

Equation~\ref{eq:WC} is exact for monoenergetic collisions. For the Brownian relative velocities, the collision efficiency factor $W_0$ takes the following form \citep{2011ApJ...731...95O}:
\begin{equation}\label{eq:WCM}
    W_0 = \exp\left( - \frac{U_{\rm C}}{k_{\rm B}T} \right).
\end{equation}
In this case, the {\it average} collision velocity is set by the Brownian velocity $v_{\rm B}=\sqrt{8k_{\rm B}T/\pi\mu}$. We use the above correction for collisions dominated by the Brownian motion ($v_{\rm B}/v_{\rm rel} > 1/2$) and Equation~\ref{eq:WC} otherwise. Such approach allows us to account for the collisional velocity distribution in the Brownian motion-dominated regime while keeping the monoenergetic recipe for the turbulence-induced collisions (as their underlying velocity dispersion depends on the turbulence nature and therefore is poorly constrained).

For sufficiently high gas ionization rates, ensuring electron abundance higher than the grain abundance, the equilibrium charge depends only on the grain radius, gas temperature $T$ and the dominant ion mass $m_{\rm ion}$:
    \begin{equation}\label{eq:Q}
        Q = \min \left\{- q(m_{\rm ion})\left(\frac{a}{0.1\,\mu \mbox{m}}\right)\left(\frac{T}{100\,\mbox{K}}\right),-1\right\}e_{\rm p},
    \end{equation}
    where $e_{\rm p}=4.8\times10^{-10}$\,statC is the proton charge and $q\approx2.3$ for HCO$^{+}$ or N$_2$H$^{+}$ ions -- see, e.g., the left panel of Figure~1 in \citet{2016ApJ...833...92I} (for $m_{\rm ion}=29$ amu). The efficient 'image' attraction \citep{1987ApJ...320..803D} ensures that even the smallest grains acquire at least one electron charge, so we set the grain charge to be at least $-e_{\rm p}$. This recipe on the grain charge can be violated in low ionization-high density regions (see Section~\ref{sec:disc}), where the limited amount of electrons reduces the charging efficiency.     This case becomes more complicated numerically and requires additional solution for the ionization and charging balance equations to account for possible dust-ion and dust-dust plasma regimes \citep{2016ApJ...833...92I}. On top of that, in the dust-ion and dust-dust plasma, the kernel of the Smoluchowski equation $K_{ij}$ starts to depend on the dust number densities $N_i, N_j$ as well, as grains compete over a limited amount of free electrons, so that their charges and, hence, collisional cross sections depend on the local number density of all surrounding grains. These effects were considered in our previous paper studying pure coagulation of charged dust \citep{2020ApJ...889...64A}. In the present paper, we add into consideration dust fragmentation, but restrict ourselves to the disk regions with high ionization fractions for simplicity. To ensure efficient dust charging and validity of Equation~\ref{eq:Q}, we apply our simulations to inner regions of protoplanetary disks, where thermal ionization of alkali elements \citep{2015ApJ...811..156D} and/or strong external ionization by the stellar X-rays allow the ionization fraction of $x_{\rm e}\gtrsim10^{-10}$. These regions are important for understanding the rocky planet formation \citep{2010A&A...515A..70D,2013A&A...556A..37D,2016ApJ...827..144F}, especially at early gravitationally unstable phases characterized by high accretion rates, luminosity outbursts, and efficient gas viscous heating \citep{2018A&A...614A..98V}.

Our implementation of dust fragmentation is similar to the approach described in \citet{2009PhDT.......360B} and \citet{2011PhDT.......279B} with several differences. First, instead of rearranging the sums in Equation \ref{eq:SCFE}, we use the quadruple precision for floating point numbers to alleviate possible rounding errors and ensure the mass conservation in the numerical scheme. The more important second modification is a correction of the actual impact velocity for the charge-charge and charge-dipole interactions between approaching grains:
\begin{equation}\label{eq:vimp}
    v_{\rm imp} = v_{\rm rel} \sqrt{1-\frac{U(a_1+a_2)}{E_{\rm kin}}},
\end{equation}
where $U$ is the electrostatic potential energy
\begin{equation}\label{eq:U}
    U(r) = \frac{Q_1Q_2}{r} - \frac{1}{2r^2}\left(\frac{\epsilon-1}{\epsilon+2}\right)\left( \frac{a_1^3Q_2^2}{r^2-a_1^2} + \frac{a_2^3Q_1^2}{r^2-a_2^2} \right)
\end{equation}
at the point of contact of two grains $r=a_1+a_2$. Here $\epsilon$ is the dielectric constant, which is $\approx7$ for silicate materials \citep{1991PCM....18....1S}. Generally, as $U(r)$ has both the repulsive Coulomb and attractive dipole components, the impact velocity can be smaller or larger than the relative velocity at infinity $v_{\rm rel}$, depending on the choice of grain sizes and charges. For physical conditions in the inner disk regions, the impact velocity is typically smaller than $v_{\rm rel}$, which softens the fragmentation conditions (see Appendix~\ref{sec:vimp}). The impact velocity is set to zero if $U(a_1+a_2)>E_{\rm kin}$. We consider Brownian motion and turbulence with Kolmogorov spectrum as dominant sources of grain relative velocities $v_{\rm rel}$. For the latter, we use analytical formulas from \citet{2021ApJ...917...82G}, which define collisional velocities for different grain size regimes (``tiny'', ``small'', and ``big'', depending on their stopping times). The turbulence is characterized by the largest eddy's size $L$ and velocity $v_{L}$, which scale with the disk vertical scale height $H$, local sound speed $c_{\rm s}$, and $\alpha_{\rm t}$-parameter as $L=\sqrt{\alpha_{\rm t}}H$ and $V_{L}=\sqrt{\alpha_{\rm t}}c_{\rm s}$ \citep{2001ApJ...546..496C}. As the turbulence-induced relative velocities between same-size grains in the ``tiny'' regime vanish \citep{2021ApJ...917...82G}, we assume that colliding grains are at least 10\% different in size. This allows us to account for the non-zero size dispersion for grains from the same mass bin \citep{2016A&A...589A..15S,2022MNRAS.514.2145B}.

\subsection{Treatment of fragmentation and erosion}

We adopt the recipe from \citet{2022ApJ...935...35S} to compute dust fragmentation and erosion. In this approach, grains experience fragmentation if their masses are similar ($\max(m_1,m_2)/\min(m_1,m_2)<10$) or erosion in the opposite case. Fragmentation produces a power-law spectrum of fragments from some minimal fragment size $a_{\rm min}$ up to the size of the largest grain among two colliding ones. In the erosion process, a smaller (projectile) grain excavates some mass from the larger one, then the sum of excavated mass (set to the mass of the projectile by default) and the projectile mass is distributed between $a_{\rm min}$ and the projectile size with a power-law spectrum. The mass of the target grain is reduced by the amount of the excavated mass and redistributed between neighbouring mass bins according to the Podolak algorithm used for coagulation treatment as well \citep{2009PhDT.......360B}. We use the power-law slope of $-3.5$ for the fragment size distribution (both for fragmentation and erosion), which translates to $-1.83$ slope in the corresponding mass distribution. The critical impact velocity required for fragmentation and erosion $v_{\rm frag}$  is assumed to be the same for both processes and all grain sizes. The fragmentation/erosion probability $(1-p_{ij})$ is set depending on the $v_{\rm imp}$ and $v_{\rm frag}$: it is zero if $v_{\rm imp}<0.8v_{\rm frag}$, unity for  $v_{\rm imp}>v_{\rm frag}$ and linearly changes with $v_{\rm imp}$ from 0 to 1 in between. We discuss potential consequences of different erosion treatment in Section~\ref{sec:disc}. The initial size distribution $f_{\rm ini}(a)$ is assumed to be a power-law with the slope $-3.5$ and maximum grain size of $0.5\,\mu$m. The minimum grain size is varied from $0.005\,\mu$m to $0.05\,\mu$m in the simulations presented below, the minimum size of fragments $a_{\rm min}$ is set to the minimal grain size in the initial distribution. To simulate higher fragmentation velocities in monomer-monomer collisions, we do not allow grain fragmentation/erosion if both grains come from the size range of the initial distribution. The canonical dust-to-gas mass ratio of $\rho_{\rm d}/\rho_{\rm g}=0.01$ and mean molecular weight of $2.3$\,amu are assumed. We work within a compact growth approximation assuming silicate grains with effective material density $\rho_{\rm s}=1.6$~g~cm$^{-3}$, which corresponds to a loose random packing of monodiperse silicate spheres with the volume fraction between $0.5-0.6$ \citep{Torquato2000}. Our simulations are local and ignore possible sedimentation and radial drift. The impact of the dust porosity and global dynamics is discussed in Section~\ref{sec:disc}. We utilize the logarithmic grain size grid with 500 bins, which covers the range from $10^{-7}$ to $2$\,cm. To integrate the system of equations~\ref{eq:SCFE} we use the semi-implicit mid-point rule for stiff systems of ODEs by \citet{BD83} implemented in \texttt{stifbs} procedure from \citet{1992nrfa.book.....P}. 

\section{Results}

   \begin{figure}
   \centering
   \includegraphics[width=\hsize]{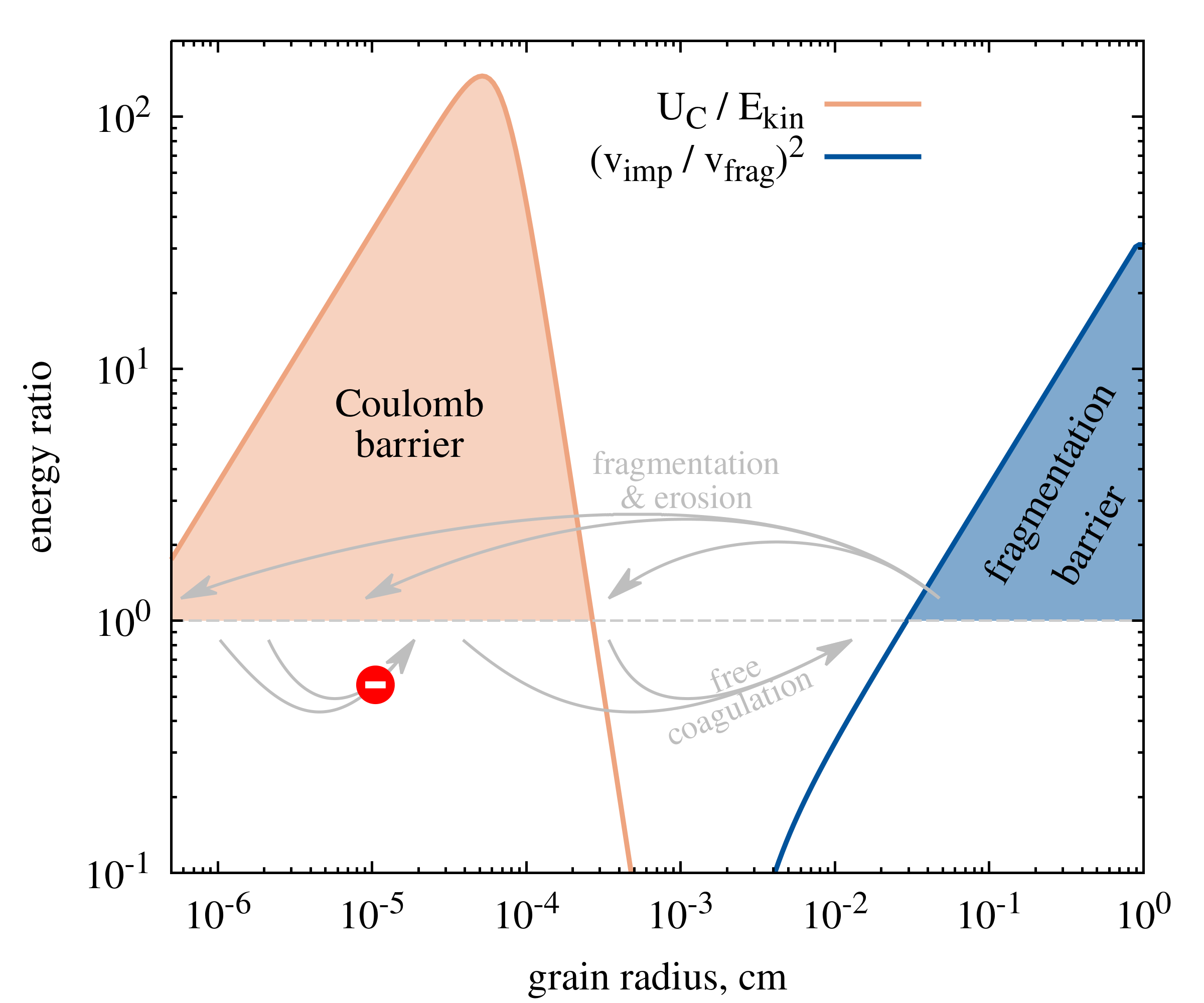}
   \caption{Electrostatic and fragmentation barriers for the collisions of similar size grains ($a_2/a_1 = 1.1$) for $\rho_{\rm g}=10^{-12}$\,g\,cm$^{-3}$, $T=1000$\,K, $\alpha_{\rm t}=3\times10^{-4}$,  $v_{\rm frag}=10$\,m\,s$^{-1}$. The impact velocity $v_{\rm imp}$ is given by Equation~\ref{eq:vimp}. Grains within the Coulomb barrier are not able to coagulate with each other, but can coagulate with grains above the barrier. Fragmentation and erosion processes repopulate the distribution at smaller sizes.}
              \label{fig:barriers}
    \end{figure}
    \begin{figure}
   \centering
   \includegraphics[width=\hsize]{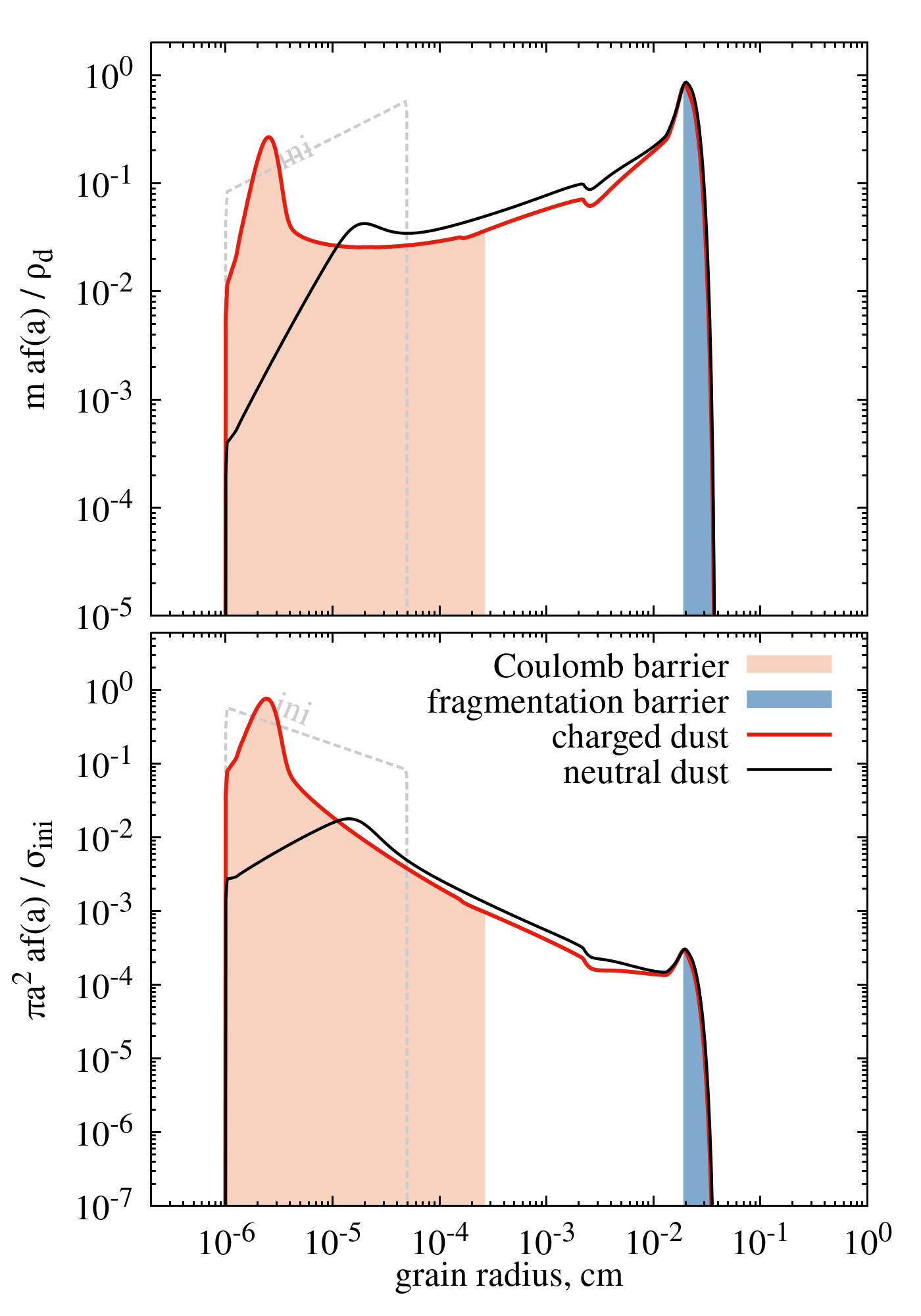}
   \caption{Dust size distribution at the coagulation-fragmentation equilibrium for neutral grains (black lines) and charged grains (red lines), computed for the conditions of Figure~\ref{fig:barriers}. The upper and lower panels display the normalized distributions of grain mass and geometrical cross section, respectively. The orange shading shows the range of grain sizes where $U_{\rm C}/E_{\rm kin}>1$ for similar-size collisions, the blue shading indicates the sizes where $v_{\rm imp}>0.8v_{\rm frag}$. The distributions in the lower panel are normalized by the initial dust cross section (per unit volume), $\sigma_{\rm ini}=6.6\times10^{-10}$\,cm$^2$\,cm$^{-3}$}.
              \label{fig:basemod}
\end{figure}
A part of the electrostatic barrier problem lies in the initial conditions. If all grains in the initial distribution fall into the size range where the Coulomb repulsion energy $U_{\rm C}=Q_1Q_2/(a_1+a_2)$ at grains' contact exceeds the initial collision kinetic energy $E_{\rm kin}=\mu_{12} v_{\rm rel}^2/2$, dust coagulation cannot start at all. Such conditions are likely met in protoplanetary disks. In Figure~\ref{fig:barriers} we show an example of typical electrostatic-to-kinetic ratio $U_{\rm C}/E_{\rm kin}$ and impact-to-fragmentation energy ratio $(v_{\rm imp}/v_{\rm frag})^2$ for similar-sized grains assuming for our reference model the gas volume density $\rho_{\rm g}=10^{-12}$\,g\,cm$^{-3}$, the temperature $T=1000$\,K, the turbulence $\alpha$-parameter $\alpha_{\rm t}=3\times10^{-4}$, and the fragmentation velocity $v_{\rm frag}=10$\,m\,s$^{-1}$ at a location $r=1$\,au around a solar mass star. The orange shading indicates the size range with electrostatically 'forbidden' coagulation: the Coulomb barrier has a maximum at sub-micron grain sizes, corresponding to a transition from Brownian to turbulence-induced motion. While dust grains from within the orange region cannot coagulate, their coagulation with larger grains is possible. Hence, the initial electrostatic barrier is easy to overcome by introducing large ($\gtrsim10\,\mu$m) seed grains, which may appear, for example, due to the radial drift from cold high-turbulence disk regions with a weak electrostatic barrier \citep{2011ApJ...731...96O}. In our simulations, we artificially allow dust to coagulate at initial stages by ignoring its charge for the first $10^{3}$\,yr, and then we switch on the charge and follow the charged dust evolution up to $10^{6}$\,yr.
 \begin{figure*}
   \centering
   \includegraphics[width=\hsize]{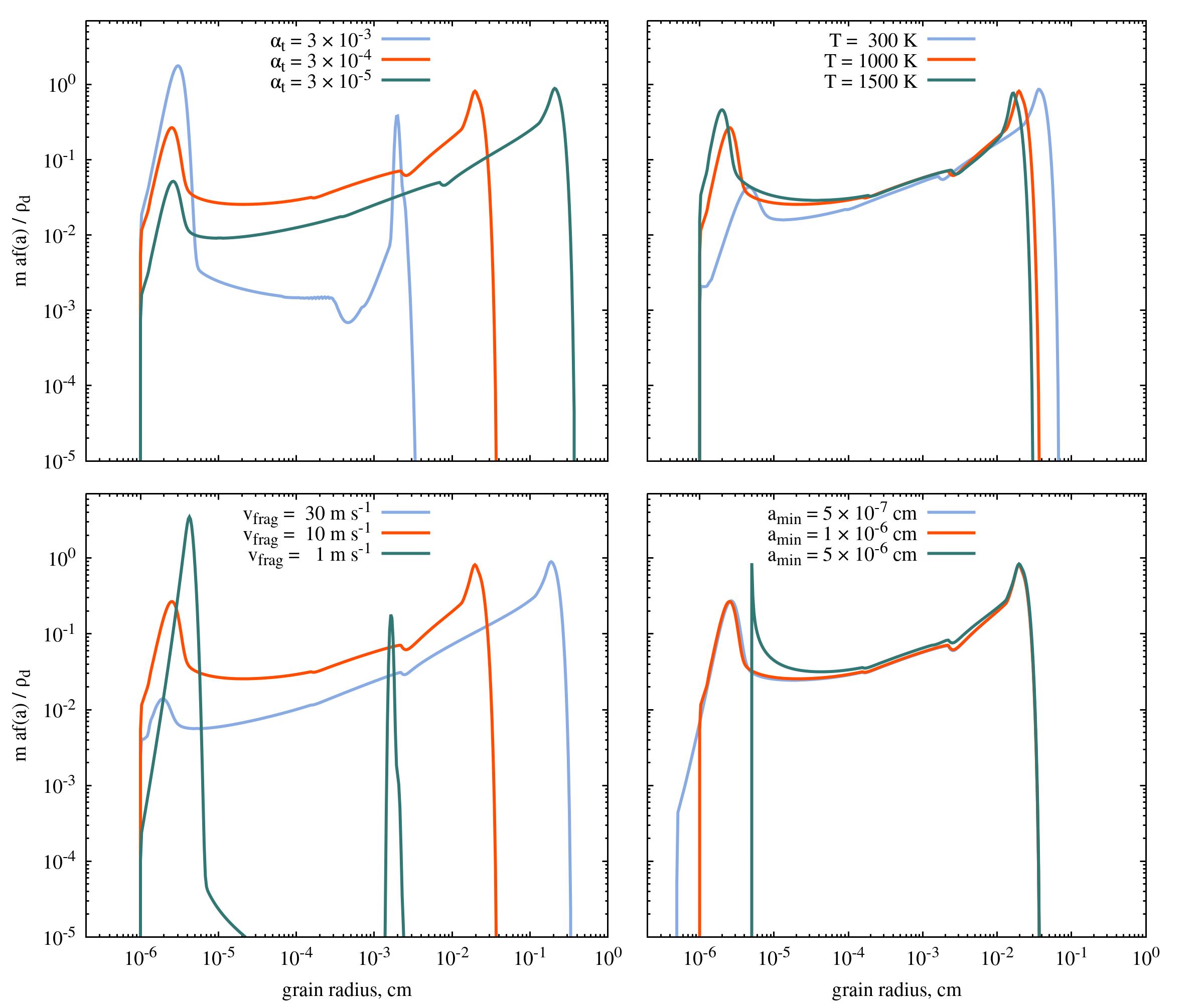}
   \caption{Dependence of the dust size distribution on the model parameters. The red lines in all panels represent the reference model plotted in Figure~\ref{fig:basemod}, the blue and green lines illustrate the effect of variation of one of the parameters, as indicated  in the legends. All distributions are for $10^6$\,yr.}
              \label{fig:parstudy}
    \end{figure*}

In Figure~\ref{fig:basemod} we demonstrate the solution of the Smoluchowski equation for the same conditions as in Figure~\ref{fig:barriers}. The upper and lower panels show the contribution of different grain sizes to the mass and total surface area, respectively. The black lines correspond to the coagulation-fragmentation equilibrium of artificially neutral grains (reached around $10^{3}$\,yr). When charging is switched on at later stages (red lines), the balance between coagulation and fragmentation changes drastically, leading to profound dust redistribution toward the smallest sizes. Due to a much larger amount of small fragments, erosion becomes more efficient. In case of neutral dust, the smallest fragments do not contribute to the erosion process significantly as they are quickly lost due to Brownian coagulation. A small kink seen in the distributions around $2\times10^{-3}$~cm reflects a transition to the regime of ``tiny'' grains in the turbulence-induced velocities \citep{2011A&A...525A..11B, 2021ApJ...917...82G}.

To study the sensitivity of the resulting size distributions to the physical conditions and model parameters, we run additional models that differ from the reference model in Figure~\ref{fig:basemod} by one of the four parameters: turbulence strength $\alpha_{\rm t}$, gas temperature $T$, fragmentation velocity $v_{\rm frag}$, and minimum size of fragments $a_{\rm min}$. The results are presented in Figure~\ref{fig:parstudy}. Some changes seen in the distributions are expected and mostly quantitative. For example, the right boundary of the size distribution depends on the fragmentation limit $a_{\rm frag}$, which scales with the physical parameters as
\begin{equation}
    a_{\rm frag} \propto  \rho_{\rm g} v_{\rm frag}^2 T^{-1/2} \alpha_{\rm t}^{-1}
\end{equation}
 (see, e.g., Eq.(34) in \citet{2016SSRv..205...41B} assuming $\rho_{\rm g} \sim \Sigma_{\rm g}/H \sim \Sigma_{\rm g}/T^{1/2}$). However, a remarkable interplay between the electrostatic and fragmentation barriers can be seen from the upper left panel: it shows a higher amount of small grains for larger values of turbulence $\alpha$-parameter -- despite the fact that stronger turbulence helps in overcoming the electrostatic barrier. While high turbulent velocities are indeed required to overcome the barrier during the initial phase of dust growth, stronger turbulence backfires when grains reach the fragmentation barrier and start to replenish the small grain ensemble. Larger values of $\alpha_{\rm t}$ suggest a lower fragmentation limit $a_{\rm frag}$ and, hence, a narrower range of the grain sizes that not affected by either barrier. These 'intermediate' grains are crucial collision partners for smaller grains under the electrostatic barrier, so their deficiency leads to a stronger pile-up of sub-micron dust. Dependence of $f(a)$ on changes in the gas density (not plotted) is very similar to that for varying $\alpha_{\rm t}$, but acting in the opposite direction in accordance with the above scaling $a_{\rm frag}\propto \rho_{\rm g}/\alpha_{\rm t}$.

 The dependencies on the gas temperature and the minimum fragment size are shown in the right panels of Figure~\ref{fig:parstudy}. Lower temperatures lead to less efficient plasma charging and, thus, to a weaker electrostatic barrier. Therefore, the number density of sub-micron grains under the barrier decreases with the temperature as well. However, one should be cautious in extrapolating this trend to the outer cold disk regions, as several important factors such as dust fractality/porosity and electron depletion are not properly considered in our simulations (see Section~\ref{sec:disc}). The dependence on the minimum fragment size shows two different shapes of the size distributions. For the small and reference values of $a_{\rm min}$ ($5\times10^{-7}$ and $1\times10^{-6}$~cm, plotted by the blue and red lines, respectively), the left peaks of the distributions are smooth, while for the large value of $a_{\rm min}=5\times10^{-6}$~cm the peak is very sharp. All three values of $a_{\rm min}$ fall into the range of a strong electrostatic barrier, where the Coulomb repulsion energy exceeding the {\it average} collision energy ($U_{\rm C}>E_{\rm kin}$). However, given the Maxwellian velocity distribution of grains, the collision efficiency factor $W_0$ calculated using Equation~\ref{eq:WCM} is non-zero and the smallest fragments can occasionally coagulate. This is not true for the large value of $a_{\rm min}=5\times10^{-6}$~cm, because the electrostatic barrier is very high in this case ($U_{\rm C}/E_{\rm kin}\sim100$). The sharpness of the peak is then enhanced by erosion, as a positive feedback loop exists between the amount of small grains and the erosion rates in the absence of efficient coagulation. Such a sharp peak is certainly a result of the used fragmentation/erosion prescription, and therefore is expected to change if more realistic size distribution models are employed for fragments (instead of the used power law distribution with fixed boundaries).
 
 Probing the characteristic range of $1-30$~m~s$^{-1}$ for the critical fragmentation velocity \citep{2008ARA&A..46...21B, 2018ApJ...853...74S} reveals a qualitatively new behavior. For the fragmentation velocity $v_{\rm frag}=1$\,m\,s$^{-1}$ (shown by the green line in the left bottom panel of Figure~\ref{fig:basemod}), we observe drastic changes, resulting in very low abundance of the grown dust by the end of simulations at $10^6$\,yr. This case represents an example of imbalanced fragmentation and erosion of the grown dust. For most of the cases shown in Figure~\ref{fig:parstudy}, the smallest fragments can collide with grown grains with impact velocities $v_{\rm imp}$ smaller than $v_{\rm frag}$, which leads to their coagulation. However, if the fragmentation velocity is low, the smallest fragments (as soon as they are able to overcome strong electrostatic barrier) collide only with $v_{\rm imp}>v_{\rm frag}$. Such collisions lead to erosion and replenish small-grain population, and therefore small fragments accumulate until the grown dust is mostly destroyed. 
 
\begin{figure}
   \centering
   \includegraphics[width=\hsize]{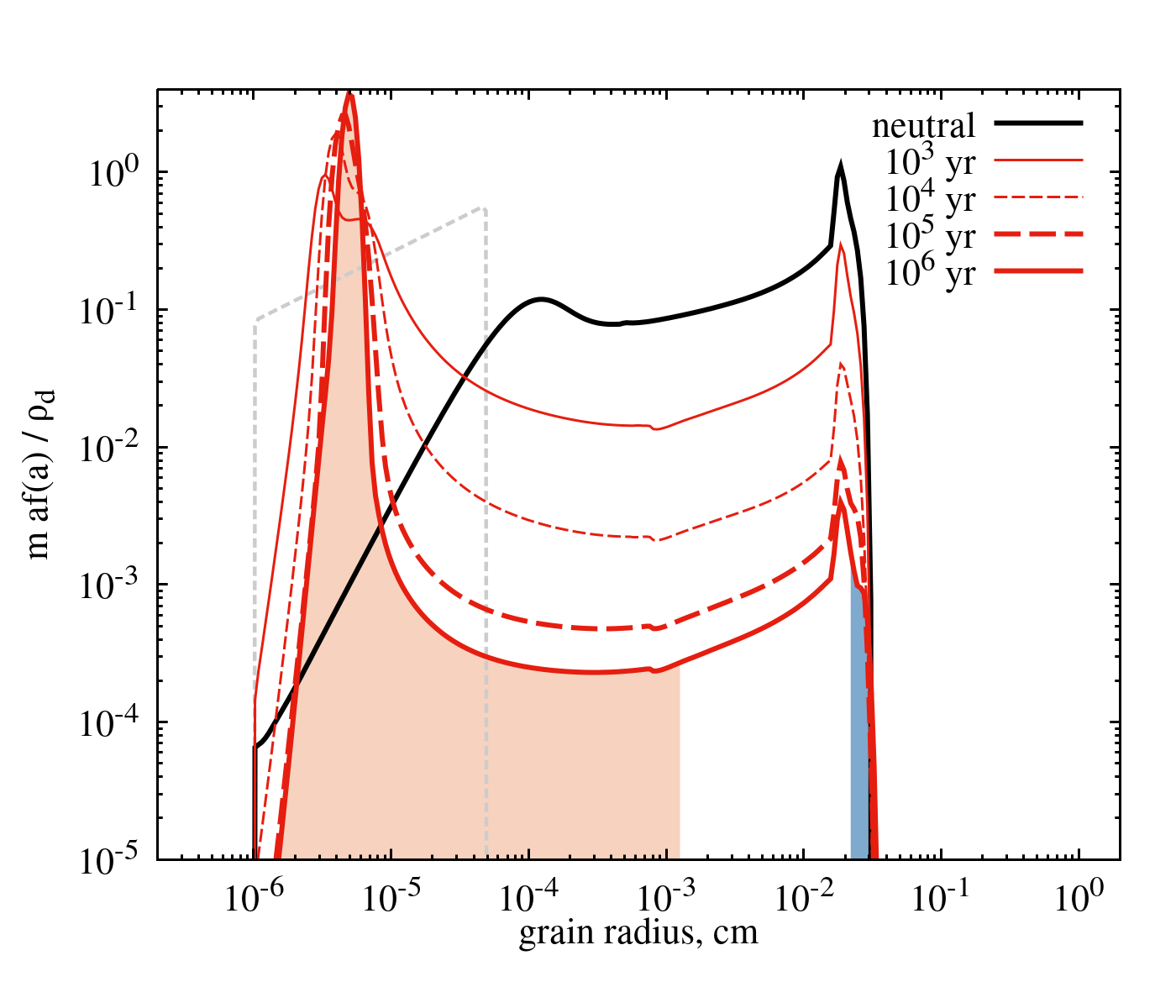}
   \caption{An example showing the size distribution of charged dust which is out of the coagulation-fragmentation equilibrium. The results are computed for $\rho_{\rm g}=5\times10^{-11}$\,g\,cm$^{-3}$, $T=1300$\,K, $\alpha_{\rm t}=3\times10^{-4}$,  $v_{\rm frag}=1$\,m\,s$^{-1}$. The black line shows the coagulation-fragmentation equilibrium for neutral dust, reached at $\sim10^3$~yr. The grain charging is switched on at this point, and then bigger grains start vanishing due to uncompensated fragmentation and erosion (as depicted by the red lines). The orange and blue shadings show the respective size ranges for the electrostatic and fragmentation barriers (assuming same-size collisions).}
              \label{fig:noneq}
    \end{figure}

 To further highlight this regime of the coagulation-fragmentation imbalance, in Figure~\ref{fig:noneq} we show how the size distribution evolves for a higher gas density of $\rho_{\rm g}=5\times10^{-11}$\,g\,cm$^{-3}$, temperature of $T=1300$\,K, and the fragmentation velocity $v_{\rm frag}=1$\,m\,s$^{-1}$. Without grain charging, the equilibrium size distribution for such density is reached within $10^{3}$\,yr (shown by the black line). Once the charging is switched on at $10^{3}$\,yr, small grains stop coagulating (see Appendix~\ref{sec:vimp}), but their replenishment due to fragmentation and erosion of grown dust continues, and vice versa -- grains with sizes within the fragmentation barrier start depleting severely, while the resulting fragments remain trapped under the electrostatic barrier. 

\subsection{Important implications}
 The presented solutions of the Smoluchowski equation suggest that the electrostatic barrier problem cannot be merely solved by introducing large seed particles, as small particles under the barrier are being constantly replenished due to fragmentation and erosion. We show that the mass contained in small grains can be comparable to or even exceed the mass of grown dust, while the surface area is completely dominated by small grains.
 
 The predicted drastic enhancement in the abundance of small grains may have profound consequences for various processes occurring in protoplanetary disks. Along with obvious implications for the grain surface chemistry, this may also significantly affect observational appearance of the disks. For example, shallow slopes of the dust millimeter opacity do not only require the presence of grown dust, but also a reduced fraction of small grains \citep[see, e.g., figure~2 in][]{2019MNRAS.486.3907P}. In the extreme case illustrated in Figure~\ref{fig:noneq}, the sharp peak of ``recycled'' grains kept under the electrostatic barrier is indistinguishable from the initial distribution when observed in the millimeter wavelengths -- since most of the grains are in the Rayleigh limit. Our analysis shows that the evolution of charged dust should typically lead to the size distributions characterized by steeper opacity indices $\beta$, as compared to the case of uncharged grains. We note that variations in the spectral index $\alpha$ are affected by several factors, which, in addition to the dust growth, also include temperature and optical depth effects \citep{2016A&A...588A..53T, 2016A&A...586A.103W, 2019ApJ...877L..22L,2021MNRAS.508.2583Z, 2022ApJ...941L..23M}, and therefore they are harder to predict.

 \section{Discussion}\label{sec:disc}
The simulations shown in the previous section reveal two key effects of dust charging on its evolution. First, the amount of sub-micron grains becomes dramatically higher than in the case of neutral grains. Second, imbalanced destruction of grown dust can occur if fragmentation velocities are $\sim1$\,m\,s$^{-1}$. At the same time, there are several poorly constrained factors and simplifying assumptions that may affect the resulting dust size distributions and shift the balance between small and grown dust in both directions. In this section, we discuss several aspects that may be important for consistent treatment of charged dust evolution in protoplanetary disks.

\subsection{Electron depletion} Equation~\ref{eq:Q} breaks down if electrons are no longer the dominant carriers of the negative charge, i.e., if most of it is carried by dust (which, according to the above results, is concentrated around $a\approx 200-500$\,\AA). For this reason, in the present paper we apply our model to the inner disk regions of $\lesssim1$\,au, where such a requirement is likely satisfied. As a rough estimate of the minimum amount of free electrons sufficient to charge the smallest grains up to at least one electron charge, one can use the condition $n_{\rm e}>n_{\rm d}$, where $n_{\rm e}$ is the number density of electrons and $n_{\rm d}$ is the dust number density (assuming the mass is dominated by grains with $a\approx a_{\rm min}$). This yields the following sufficient condition for the minimum value of the ionization fraction $x_{\rm e}=n_{\rm e}/n_{\rm g}$:  
\begin{equation}
\begin{aligned} 
    x_{\rm e} &> 2\times10^{-10}\left(\frac{a_{\rm min}}{300\,\mbox{\AA}}\right)^{-3}\left(\frac{\rho_{\rm d}/\rho_{\rm g}}{0.01}\right)\\
    &\times\left(\frac{\rho_{\rm s}}{1.6\,\mbox{g cm}^{-1}}\right)^{-1}\left(\frac{\mu_{\rm g}}{2.3 \mbox{ amu}}\right),
\end{aligned}
\end{equation}
which is typically satisfied for the inner disk regions, where the electron abundance is governed by efficient thermal ionization of alkali elements \citep{2015ApJ...811..156D}.

Electrons in the outer, low-ionization regions of protoplanetary disks can be strongly depleted if a large amount of small grains is present \citep{2016ApJ...833...92I}. In such cases, the smallest $\lesssim 100\,\mbox{\AA}$ fragments can become neutral and therefore free to coagulate until the resulting grains become large enough to acquire (at least) one electron charge. Due to a strong dependence of the depletion condition on the grain size \citep[see Equations 16 and 27 in][]{2016ApJ...833...92I}, this is likely to occur for the sizes within the electrostatic barrier. This suggestion is supported by our earlier modelling of charged dust coagulation in the presence of electron depletion \citep{2020ApJ...889...64A}, demonstrating that the electrostatic barrier efficiently limits the dust growth under conditions of the MRI-dead zones. We expect that accounting for the lower charging efficiency in the outer disk regions would shift the left peak of dust distribution to larger sizes, still keeping large amounts of small fragments under the electrostatic barrier.

\subsection{Erosion efficiency}
 In our simulations, the critical velocity $v_{\rm er}$ above which the erosion occurs is set equal to $v_{\rm frag}$. In fact, the erosion experiments for the projectile sizes $a_{\rm p}$ between $1$ and $100\,\mu$m \citep{2018ApJ...853...74S} show that the critical erosion velocity scales as $v_{\rm er}\sim a_{\rm p}^{0.62}$, i.e., erosion is achieved easier for smaller projectiles. Also, the erosion efficiency (the ratio of the small fragments mass to the initial mass of the projectile) is set equal to 2 for any collision that leads to erosion, with no dependence on $v_{\rm imp}$. However, experimental and theoretical studies suggest that the erosion efficiency linearly increases with $v_{\rm imp}$ \citep{2018ApJ...853...74S}, and therefore more energetic collisions excavate more mass. Nevertheless, with both factors underestimating the erosion efficiency, we still obtain anomalously high abundance of small grains.
 
 Details of erosion are less important for models ignoring grain charging because of fast coagulation of small grains and, hence, their low abundance. While it is unknown whether the scaling relations for the erosion efficiency and critical velocity hold also for sub-microns grains, accounting for the two above factors will likely lead to higher amounts of small dust and relaxed conditions for the imbalanced fragmentation.
        
 \subsection{Dust porosity} The very possibility for overcoming the electrostatic barrier relies on a sufficient strength of non-thermal collision velocities of micron-size grains. Small porous grains are better coupled to the gas than compact grains of the same mass, i.e., the former have lower turbulence-induced velocities. Thus, dust porosity hampers charged dust coagulation and widens the mass range of grains affected by the electrostatic barrier \citep{2011ApJ...731...95O, 2020ApJ...889...64A}. Our simulations are done within the simplifying compact growth approximation, so they underestimate the strength of the electrostatic barrier. 
    
 \subsection{Turbulence spectrum} Our modelling assumes the Kolmogorov turbulence with the slope of the kinetic energy spectrum of $p=-5/3$. Shallower spectra for the Iroshnikov--Kraichnan cascade \citep[$p=-3/2$, see ][]{1964SvA.....7..566I, 1965PhFl....8.1385K}, or for the turbulence seen in magneto-hydrodynamical simulations \citep[$p=-4/3$, see][]{2021ApJ...909..148G, 2020ApJ...891..172G} are more favorable for early coagulation and may help to overcome the electrostatic barrier \citep{2021ApJ...917...82G}. On the other hand, within a certain range of turbulent scales the slopes for pure hydrodynamic instabilities (such as the subcritical baroclinic instability or the vertical shear instability) are steeper than the Kolmogorov one \citep{2003ApJ...582..869K, 2020MNRAS.499.1841M}. Thus, deviations from the Kolmogorov turbulence can shift the balance between small and grown dust in both directions.
    
 \subsection{Space or time-dependent turbulence strength} Overcoming the electrostatic barrier requires strong turbulence and/or presence of grown grains. Avoiding dust destruction at the fragmentation barrier and quenching the replenishment of small dust under the electrostatic barrier requires the opposite. A viable way to overcome both barriers may lie in dealing with them subsequently and not simultaneously. Grain coagulation in high-turbulence outer disk regions and their subsequent drift to dead zones is one perspective way to be explored \citep{2011ApJ...731...96O}. Another possibility is time-dependent non-thermal sources of grain collisional velocities. Early gravitationally unstable disk phases may provide conditions to overcome the electrostatic barrier and to form certain amounts of grown dust. Subsequent global disk evolution towards less turbulent state may help to alleviate the fragmentation barrier and drag the remaining small grains towards next steps in bottom-up planet formation scenario. These possibilities are worth to be studied in the future with the next generation dust evolution models taking into account global disk evolution \citep{2019MNRAS.490.4428P, 2020ApJ...889L...8L, 2020MNRAS.499.5578A, 2021MNRAS.507.2318V, 2022MNRAS.514.2145B}.

\section{Conclusions}
Electrostatic interaction between dust grains is a frequently overlooked factor in contemporary models of dust evolution in protoplanetary disks. The Coulomb repulsion of same-charged grains is likely to be an effective barrier blocking the initial dust coagulation of micron-size grains \citep{2009ApJ...698.1122O, 2011ApJ...731...96O}. One possibility to trigger the initial coagulation is to introduce large ($\gtrsim10\,\mu$m) seed particles, which can coagulate with smaller particles and thus pull them out of the size range corresponding to the electrostatic barrier. In this paper, we perform a combined analysis of electrostatic interactions between dust grains and their fragmentation, and develop a self-consistent approach to model these two processes. For this purpose, we numerically solve the Smoluchowki coagulation-fragmentation equation for a set of physical conditions relevant for inner regions of protoplanetary disks with ionization fractions $\gtrsim10^{-10}$. Our conclusions can be summarized as follows:
   \begin{enumerate}
      \item Coagulation-fragmentation equilibrium for charged dust is characterized by dramatically higher abundances of sub-micron particles than in the case of neutral dust. The reason for that is a continuous replenishment of small grain population due to fragmentation of macroscopic grains. Without charging, the sub-micron grains quickly coagulate owing to their Brownian motion, which is not the case if charging is taken into account. 
      \item For studied cases with the fragmentation velocities $v_{\rm frag}\sim 1$~m~s$^{-1}$, the coagulation-fragmentation balance between smaller and larger grains cannot be reached. This is because the electrostatic barrier inhibits almost all low-velocity collisions, that may lead to coagulation, while the majority of high-velocity collisions, able to overcome the barrier, are destructive for sufficiently small $v_{\rm frag}$. The resulting lack of mass supply to larger grains causes their gradual destruction on timescales shorter than the typical disk lifetimes.
   \end{enumerate}
Further studies of charged dust evolution, focusing on the outer disk regions and searching for ways to overcome the combined effect of electrostatic and fragmentation barriers, may include a more detailed analysis of erosion and charging efficiencies, dust porosity, turbulence properties, and global disk evolution.

\begin{acknowledgments}
We thank anonymous referee for their constructive suggestions. VA was supported by the RSF grant 22-72-10029.
\end{acknowledgments}

\appendix

\section{Effect of electrostatic interaction on collisional cross sections}\label{sec:inddipcros}
The collision cross section between two charged grains cannot be written analytically if both repulsive (Coulomb) and attractive (induced dipole) terms are present. Here we describe the algorithm used to find the collision efficiency factor $W$ for the cross section numerically:
\begin{equation}
    \sigma=\pi(a_1+a_2)^2W.
\end{equation}
The electrostatic potential energy as a function of the radial distance $r$ between two spherical particles is (\citet[][$\S3$]{1960ecm..book.....L}; \citet{1987ApJ...320..803D}):
\begin{equation}
    U(r) = \frac{Q_1Q_2}{r} - \frac{1}{2r^2}\left(\frac{\epsilon-1}{\epsilon+2}\right)\left( \frac{a_1^3Q_2^2}{r^2-a_1^2} + \frac{a_2^3Q_1^2}{r^2-a_2^2} \right).
\end{equation}
This function is non-monotonic: the Coulomb repulsion of same charged dust grains, dominating at larger distances, changes to attraction at smaller distances due to the induced dipole terms. We assume a conservative value of the dielectric constant $\epsilon=7$ for silicate materials \citep{1991PCM....18....1S}. The dynamics of two colliding grains is described by the effective potential energy, which takes into account the centrifugal component at non-zero impact parameter~$b$ \citep[][$\S14$]{1969mech.book.....L}:
\begin{equation}
    U_{\rm eff}(r,b) = U(r) + K_{\infty}\left(\frac{b}{r}\right)^2,
\end{equation}
where $K_{\infty}=\mu_{12} v_{\rm rel}^2/2$ is the kinetic energy of the collision at infinity. 

The algorithm to calculate the cross section consists of two steps. Step~1 is required to determine if collisions are possible for zero impact parameter; if yes, then the collision cross section is calculated at step~2.

1. Finding the location $r_{\rm max}$ of the maximum of $U(r)$ for $r\geq a_1+a_2$. If the dipole component is sufficiently strong, the maximum is located at $r_{\rm max}>a_1+a_2$, derived from the condition
\begin{equation}
   \frac{dU}{dr} = 0.
\end{equation}
Otherwise, if the derivative is negative at $r\ge a_1+a_2$, the maximum is achieved at $r_{\rm max}=a_1+a_2$. If $U(r_{\rm max})>K_{\infty}$, then grains cannot collide and $W=0$.

2.  Calculating the cross section. If $U(r_{\rm max})\leq K_{\infty}$, there is a critical impact parameter $b_*$ for which $U_{\rm eff}(r,b_*)=K_{\infty}$, i.e., for $b<b_*$ grains overcome the potential barrier and collide. Thus, the collision cross section is equal to $\pi b_*^2$, and  
\begin{equation}\label{eq:WCD}
    W=\frac{b_*^2}{(a_1+a_2)^2}.
\end{equation}
The desired value of $b_*$ is a root of the equation
\begin{equation}
    U_{\rm eff}(r_{\rm max}(b),b) = K_{\infty},
\end{equation}
where $r_{\rm max}(b)$ is, in turn, a root of 
\begin{equation}\label{eq:dUeffdr}
   \frac{\partial U_{\rm eff}(r,b)}{\partial r} = 0.
\end{equation}
For a given $b$, if the derivative is negative at $r\ge a_1+a_2$, then $r_{\rm max}(b)= a_1+a_2$.

The above algorithm is formulated for like-charged grains, i.e., for $Q_1Q_2>0$. If one of the grains is neutral, the interaction is determined by the attractive dipole term and the potential energy $U(r)$ is negative for any $r$. In this case, step~1 can be skipped and step~2 yields $W\ge1$. We note that the analysis becomes more complicated for $Q_1Q_2<0$, as Equation~\ref{eq:dUeffdr} may then have two roots, in which case $W$ should be computed for the root corresponding to a maximum. However, the latter regime is usually realized in the upper atmosphere of disks and therefore is irrelevant for our analysis.

\begin{figure}
   \centering
   \includegraphics[width=0.52\hsize]{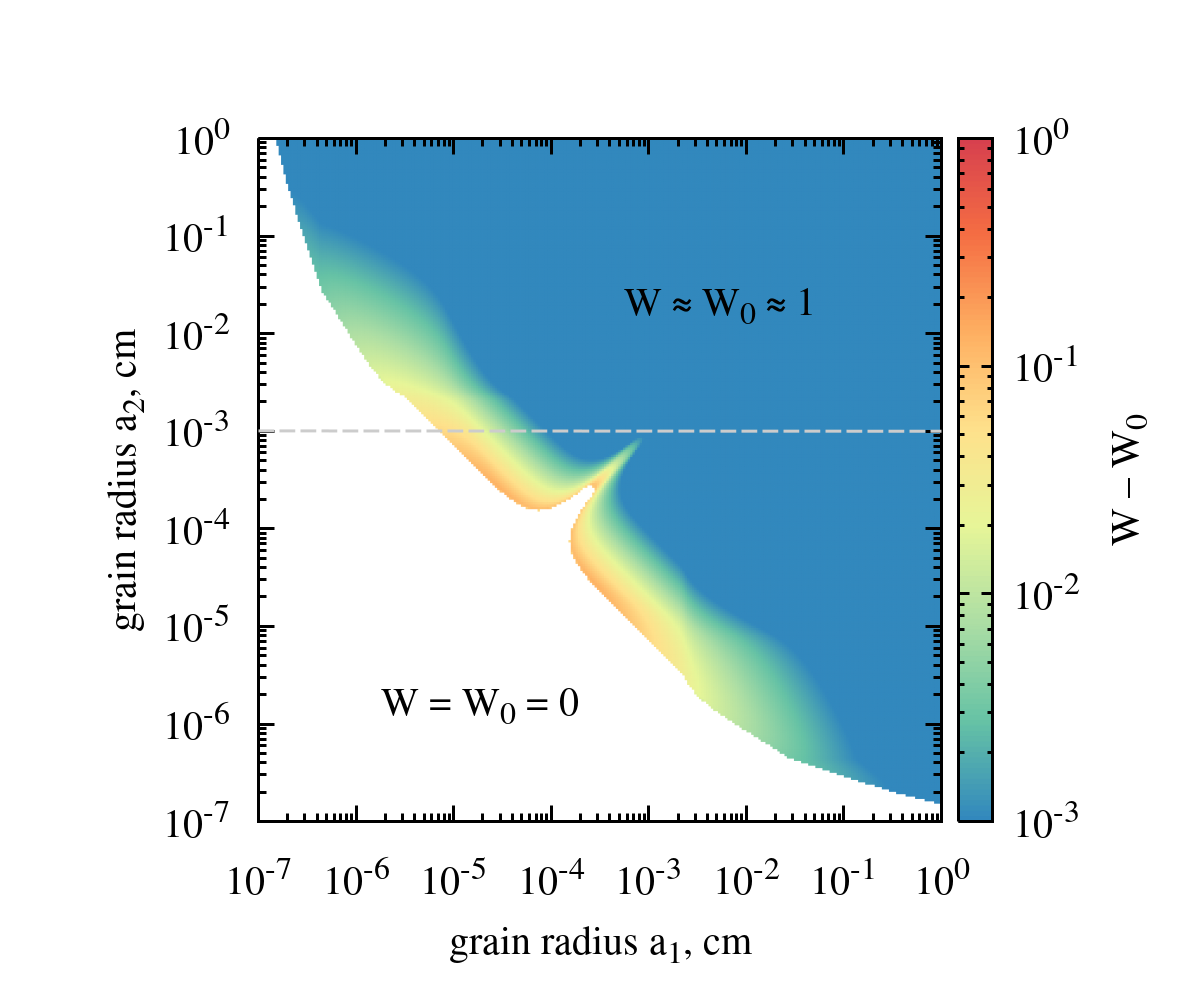}
   \includegraphics[width=0.47\hsize]{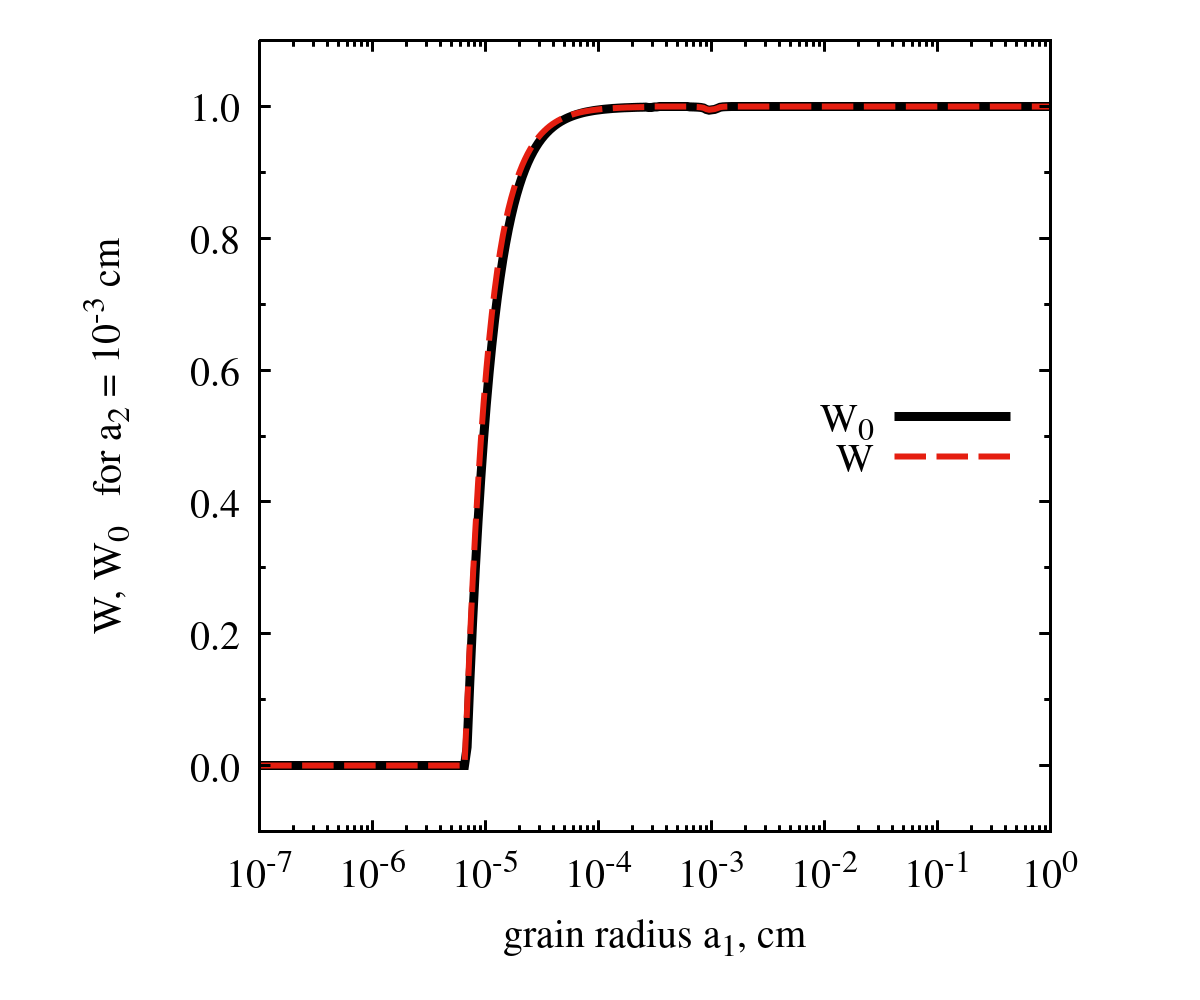}
   \caption{Left panel: difference between the collision efficiency factor $W$ calculated with the dipole attraction taken into account (Equation~\ref{eq:WCD}), and without it, $W_0$ (where only the Coulomb repulsion is included, Equation~\ref{eq:WC}). The white area indicates the total electrostatic blocking of collisions. Right panel: the values of $W_0$ and $W$ for $a_2=10^{-3}$\,cm (corresponds to the dashed line in the left panel). The physical conditions are the same as in Figure \ref{fig:barriers}.}
              \label{fig:WWC}
\end{figure}

In Figure~\ref{fig:WWC} we show the difference between the pure Coulomb collision efficiency factor $W_0$, calculated analytically using Equation~\ref{eq:WC}, and the full factor $W$ with dipole terms, calculated with the above algorithm. The physical parameters correspond to our reference model. The collision efficiency factor with dipole terms $W$ is always larger or equal to that with the Coulomb term only, $W_0$. The difference between numerical and analytical approaches becomes noticeable at intermediate values of $W$, reaching the maximum of $\approx 0.15$. However, such a difference is observed in a relatively narrow range of grain sizes and has marginal impact on the resulting dust size distributions. In the present paper we utilize the analytical approach, as it allows a simple recipe to account for the grain velocity dispersion (Equations \ref{eq:WC} and \ref{eq:WCM}).

\section{Effect of electrostatic interaction on impact velocities}\label{sec:vimp}
The relative velocity between two charged grains changes as they approach each other. As the electrostatic interaction includes both the long-range Coulomb repulsion and the short-range induced-dipole attraction (Equation~\ref{eq:U}), the actual impact velocity at the moment of grain collision $v_{\rm imp}$ (Equation~\ref{eq:vimp}) can be smaller or larger than the initial relative velocity at infinity $v_{\rm rel}$. The ratio $v_{\rm imp}/v_{\rm rel}$ depends on the grain charges and sizes. For the inner disk conditions considered in this paper, the Coulomb term dominates and $v_{\rm imp}$ for charged grains is generally smaller than that for neutral grains of the same sizes. However, the opposite situation is possible for outer (colder) disk regions.

\begin{figure}
   \centering
   \includegraphics[width=0.45\hsize]{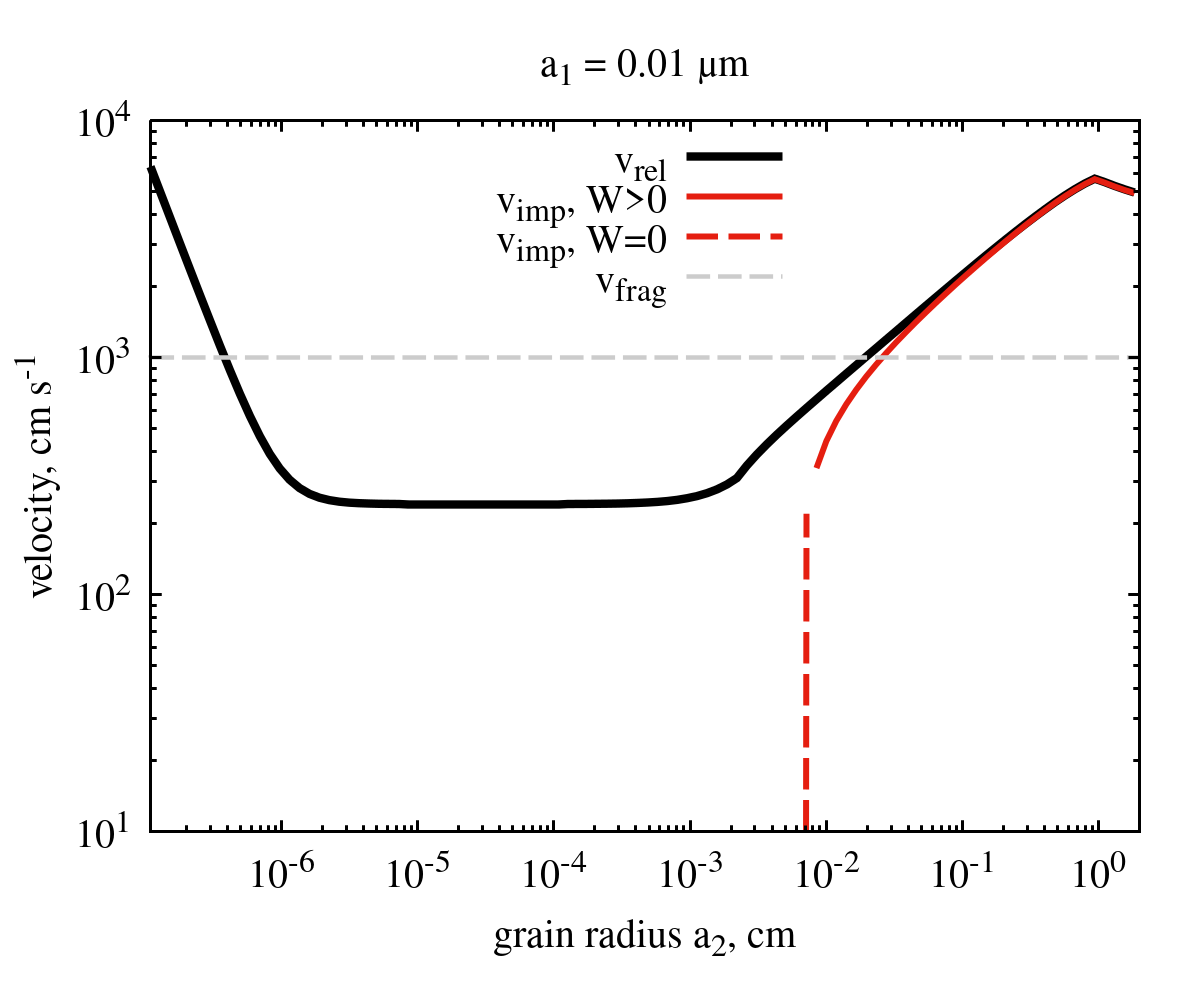}
   \includegraphics[width=0.45\hsize]{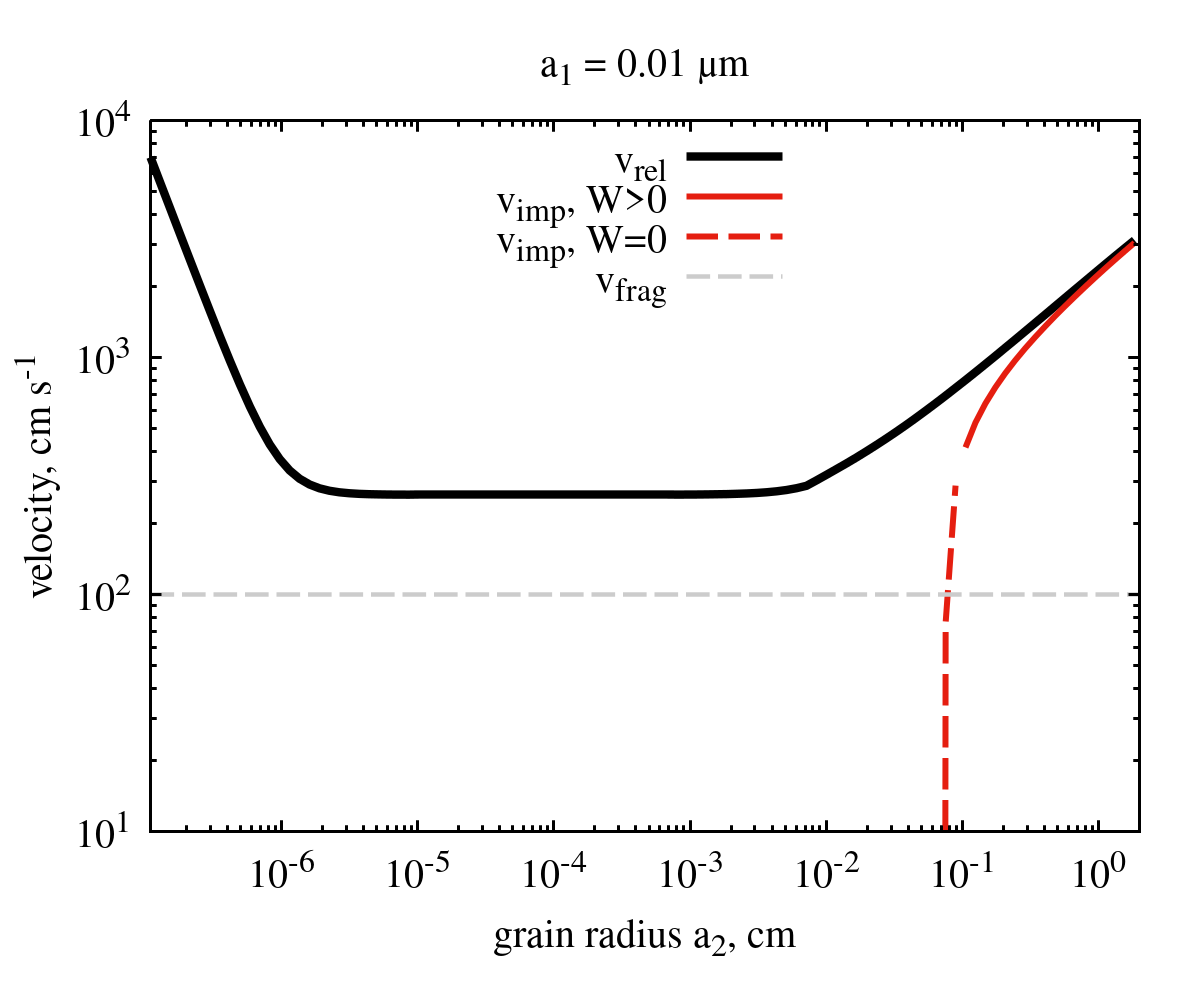}
   \caption{Relative velocity at infinity $v_{\rm rel}$ (black line) and upon collision $v_{\rm imp}$ (red lines) between the smallest fragments $a_1=0.01\,\mu$m and other grains. The red lines become dashed when the collision cross section factor $W$ is zero. Left panel: The reference model (Figures \ref{fig:barriers} and \ref{fig:basemod}). Right panel: the case of reduced fragmentation velocity (Figure~\ref{fig:noneq}).}
              \label{fig:vrelimp}
\end{figure}

In Figure~\ref{fig:vrelimp} we show how the electrostatic interaction affects the impact velocities between small grains with $a_1=0.01\,\mu$m and other grains. The left panel corresponds to our reference model with $v_{\rm frag}=10$\,m\,s$^{-1}$ (Figures \ref{fig:barriers} and \ref{fig:basemod}), while the right panel is for the model with lower fragmentation velocity of $1$\,m\,s$^{-1}$ (Figure~\ref{fig:noneq}), where the grown grains are being efficiently destroyed. The solid red lines showing $v_{\rm imp}$ change to the dashed where the collision efficiency $W$ becomes equal to zero. As a result, no coagulation-fragmentation equilibrium is reached for the case depicted in the right panel, as the smallest fragments can only collide with $v_{\rm imp}>v_{\rm frag}$. 
\bibliography{sample631}{}
\bibliographystyle{aasjournal}

\end{document}